\documentclass[preprint]{elsarticle}

\usepackage{hyperref}

\journal{Nucl. Instrum. Meth. Phys. Res. A}

\begin{document}

\begin{frontmatter}

\title{A novel measurement approach for near-edge x-ray absorption fine structure: continuous 2$\pi$ angular rotation of linear polarization}

\author[label1]{Yoshiki Kudo}
\address[label1]{Institute for Solid State Physics, The University of Tokyo, Kashiwa, Chiba 277-8581, Japan}

\author[label2]{Yasuyuki Hirata}
\address[label2]{Department of Applied Physics, National Defense Academy, Japan}

\author[label1]{Masafumi Horio}

\author[label1,label3]{Masahito Niibe}
\address[label3]{Laboratory of Advanced Science and Technology for Industry, University of Hyogo, 3-1-2 Koto, Kamigori-cho, Ako-gun, Hyogo, 678-1205, Japan}

\author[label1,label4]{Iwao Matsuda\corref{mycorrespondingauthor}}
\address[label4]{Trans-scale Quantum Science Institute, University of Tokyo, Bunkyo-ku, Tokyo 113-0033, Japan}



\cortext[mycorrespondingauthor]{Corresponding author}
\ead{imatsuda@issp.u-tokyo.ac.jp}
\ead{TEL:+81-(0)4-7136-3402}


\begin{abstract}
A new technical method is developed for soft X-ray spectroscopy of near-edge x-ray absorption fine structure (NEXAFS). The measurement is performed with continuously rotating linearly polarized light over 2$\pi$, generated by a segmented undulator. A demonstration of the rotational NEXAFS experiment was successfully made with a 2D film, showing detailed polarization-dependence in intensity of the molecular orbitals. The present approach provides varieties of technical opportunities that are compatible with the state-of-the-art experiments in nano-space and under the $operando$ condition.   
\end{abstract}

\begin{keyword}
polarization, soft X-ray, undulator, absorption spectroscopy
\end{keyword}

\end{frontmatter}


\section{Introduction}
Knowledge of the chemical or electronic state of a material has been significant in various fields of science and technology. An X-ray beam is the well-known probe and it has been widely used at synchrotron radiation facilities over the world. Near-edge x-ray absorption fine structure (NEXAFS) is one of the X-ray spectroscopy techniques that have been adopted for electronic analysis of a sample\cite{Stohr,Zhang2020,Kobl2020,Tateishi,Yamamoto2019,Auwarter2018,Niibe2018}. Recently, NEXAFS measurements have been developed to the nano-space imaging and the $operando$ experiments that allows us to optimize functionalities of molecular nanodevices, for example\cite{Zajac2021,Watt2012,Fukidome2014}. Using linearly-polarized light, the method has benefit to specify the orbital-type of a state or to determine the molecular configuration\cite{Stohr}. In such measurements, NEXAFS spectra are taken at various angles between molecule and linear polarization. At a typical beamline of the synchrotron radiation facility, the linear polarization of the beam is fixed and a sample is rotated to collect a series of the angle-resolved data. The experimental approach is concise but, at the same time, the rotation scheme restricts conditions of the sample environment. For example, a sample should be uniform to cover a possible misalignment between the beam spot and the rotation axis of a sample. On the other hand, in a case of the nano-beam NEXAFS or $operando$ experiment, it is technically too challenging to precisely secure the beam position on a non-uniform and complicated nanodevice after the sample rotation. Thus, it has become demanded to rotate the linear polarization of X-ray beam, instead of a sample, to meet requirements of the state-of-the-art experiment today.

A polarization control of the X-ray beam technically depends on the photon energy range, hard and soft X-ray regions. While the light polarization can be regulated by an optical element, $i.e.$ a diamond phase retarder, for hard X-ray, it is only possible by a polarization control undulator, such as an APPLE-type or segmented-type, for soft X-ray\cite{Sasaki1993, Sasaki1994, Yamamoto2014,Matsuda2019}. The APPLE (APPLE-II) undulator controls the polarization by changing the electron trajectory by mechanical arrangements of the magnet arrays. This type of the light source has been widely installed in various synchrotron radiation facilities and the linear polarizations are typically used at two angles that are perpendicular to each other, horizontal and vertical. In a case of the segmented undulator, it makes the polarization control by regulating a phase shift of electromagnetic waves that are generated in the undulator segments \cite{Tanaka2002, Tanaka2004}. This type of the undulator is novel and it is successfully installed at SPring-8 BL07LSU\cite{Yamamoto2014,Senba2011}. The continuous variation of the phase shifts has realized smooth polarization controls of soft X-ray beam between linear and circular ones with fast switching (13 Hz). The new regulation has achieved to determine the element-specific complex permittivity in soft X-ray\cite{Kubota2017}.

In the present research, we designed a new control of linear polarization at any angle with this segmented undulator. We generated soft X-ray beam with linear polarization that continuously rotates its angle over 2$\pi$ and we measured NEXAFS spectra of a 2D material, showing clear angle-dependence with the molecular orbitals. The new experimental method of NEXAFS, rotational NEXAFS, provides varieties of technical opportunities that are compatible with experiments in nano-space and under the $operando$ condition.   

\begin{figure*}[htp]
 \begin{center}
 \includegraphics[width=12cm]{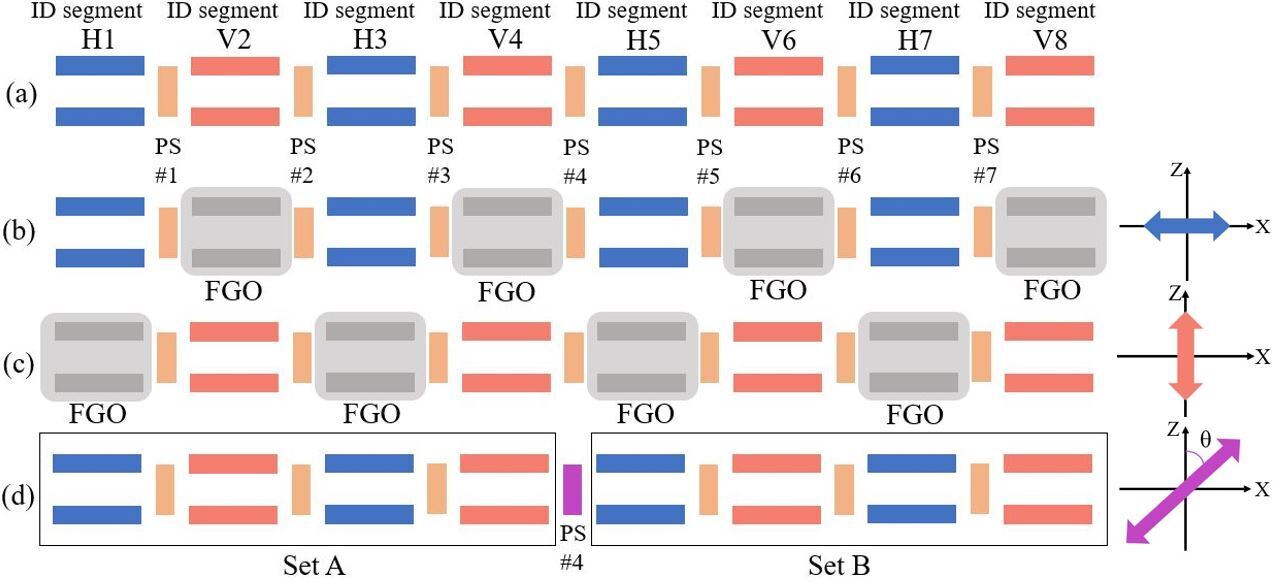}
   \caption{ Schematic drawing of the segmented cross undulator and its generations of linearly polarized light. (a) Components of the light source: eight segments of insertion device (ID) and seven phase shifters (PS's). The horizontal (H) and vertical (V) ID segments are arranged alternatively. The PSs are placed between the two segments. ID and PS are labeled with numbers. (b) Horizontally linearly polarized radiation, generated by the H1, H3, H5, and H7 segments.  (c) Vertically linearly polarized radiation, generated by the V2, V4, V6, and V8 segments. (d) Rotational linearly polarized radiation, developed by a combination of two circular polarized light that are generated at Set A and Set B. A phase shift at PS$\sharp$4 determines an angle, $\theta$, of the linear polarization. In the figure, FGO indicates the full-gap opening of an ID segment.} 
  \label{fig_1} 
 \end{center}  
\end{figure*}
	
\section{Experimental}
Continuous polarization modulation of a soft X-ray beam was made using a segmented cross undulator at SPring-8 BL07LSU\cite{Yamamoto2014}. The light source is composed of eight segments of insertion device (ID) and seven phase shifters (PSs), as shown in Fig.1. The ID segment is a horizontal  figure-8 undulator or a vertical figure-8 undulator. The horizontal (H) and vertical (V) ID segment are placed alternatively and sandwich a PS. A PS is used to adjust the relative phase of radiation emitted from each segment by changing the path length of the electron orbit in magnetic field. In the present experiment, the path was controlled by a distance between permanent magnets in a PS. Four horizontal ID segments, H1, H3, H5, and H7, generate horizontally linearly polarized radiation at the fundamental radiation and the other four vertical ID segments generate vertically linearly polarized radiation. The PS components are used to optimize the optical interference of undulator radiation, generated at individual segments. Horizontally linearly polarized light is produced by the H1, H3, H5, and H7 segments [Fig.1 (b)], while vertically linearly polarized by the V2, V4, V6, and V8 ones [Fig.1 (c)]. To realize linearly polarized light with arbitrary angle, $\theta$, we combined two waves of the left- and right-handed circularly polarized light, generated by sets of the ID segments, Set A and Set B, as shown in Fig.1(d). Set A is composed of the H1, V2, H3, and V4 segments, while Set B is assembled with the H5, V6, H7, and V8 segments. When Set A produces right-handed (left-handed) circular polarized light, we made Set B generate left-handed (right-handed) circular polarized light. Then, the rotation angle ($\theta$) of linearly polarized light was controlled by changing the phase shift between the left and right circularly polarized light at PS$\sharp$4.

For demonstration of an NEXAFS experiment with the novel polarization control, the experiment was made in an ultrahigh vacuum chamber at the beamline SPring-8 BL07LSU. We prepared a film of the hexagonal boron nitride (h-BN) as a sample\cite{Auwarter2018,Niibe2018}. Microcrystals of h-BN has a flat shape and the film becomes two-dimensionally oriented when mechanically pressed\cite{Niibe2018}. Prior to the UHV installation, powder of h-BN was placed on an indium sheet on a copper plate and compressed with a hand presser. NEXAFS spectra were recorded by total electron yield at room temperature.

\section{Demonstration of rotational NEXAFS}
Figure 2 collects N K-edge NEXAFS spectra of the oriented h-BN film, measured with (a) horizontally or (b) vertically linearly polarized light. We set the horizontal (vertical) direction and axis of the sample rotation are perpendicular (parallel) to each other. As shown in the figure, two prominent peaks are found at hv = 401.5 and 408.0 eV that correspond to the $\pi^{*}$ and $\sigma^{*}$ orbitals of h-BN\cite{Auwarter2018,Niibe2018}. In the experiment of the horizontal polarization [Fig. 2(a)], intensity of the $\pi^{*}$ peak is larger in the grazing incidence geometry (incident angle of 80$^{\circ}$) than in the normal incidence (incident angle of 0$^{\circ}$). On the other hand, the appearance is opposite for the $\sigma^{*}$ peak. This spectral behavior is described in terms of the dipole transition in light absorption\cite{Stohr}. Since the $\pi^{*}$ ($\sigma^{*}$) orbital is oriented along the out-of-plane (in-plane), the peak intensity enhances when linear polarization is parallel to the out-of-plane (in-plane). The consistent results are found in a case of the vertical polarization [Fig. 2(b)]. In a case of the grazing and normal incidence, linear polarization is always parallel to the in-plane. Thus, in these measurement geometries, only a peak of the $\sigma^{*}$ orbital appears prominently.

\begin{figure}[htp]
 \begin{center}
  \includegraphics[width=8cm]{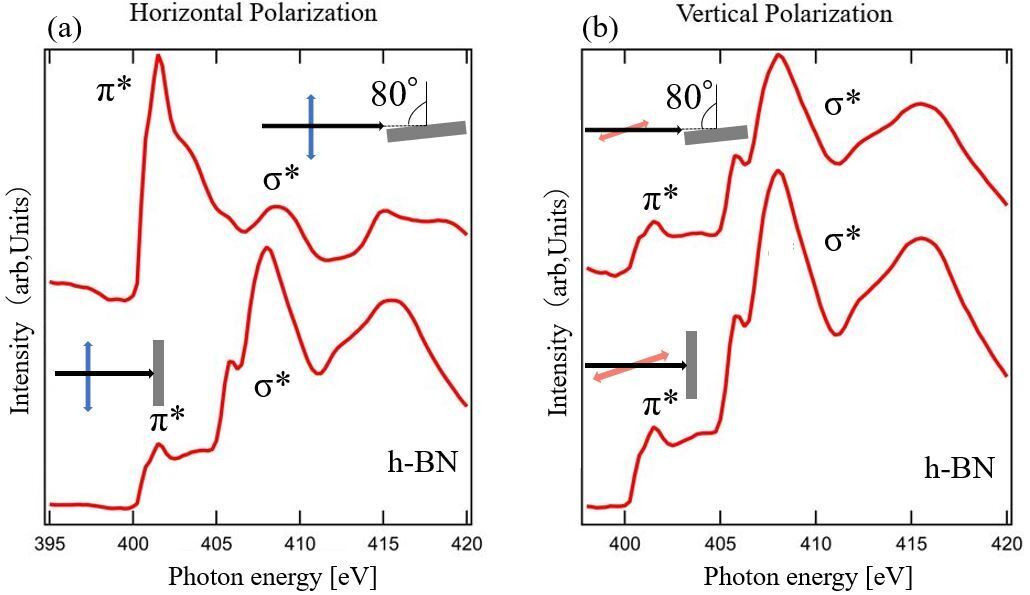}
   \caption{A collection of N K-edge NEXAFS spectra of the oriented h-BN film, measured with soft X-ray of the (a) horizontally and (b) vertically linear polarization. The set A and set B are chosen to be right- and left-handed circularly polarized light, respectively. The incident angle is referred from the surface normal. The grazing and normal incidence correspond to incident angles of 80$^{\circ}$ and 0$^{\circ}$, respectively. Measurement geometries of individual spectra are indicated in the figure. } 
  \label{fig_2} 
 \end{center}  
\end{figure}
\begin{figure}[htp]
 \begin{center}
  \includegraphics[width=8cm]{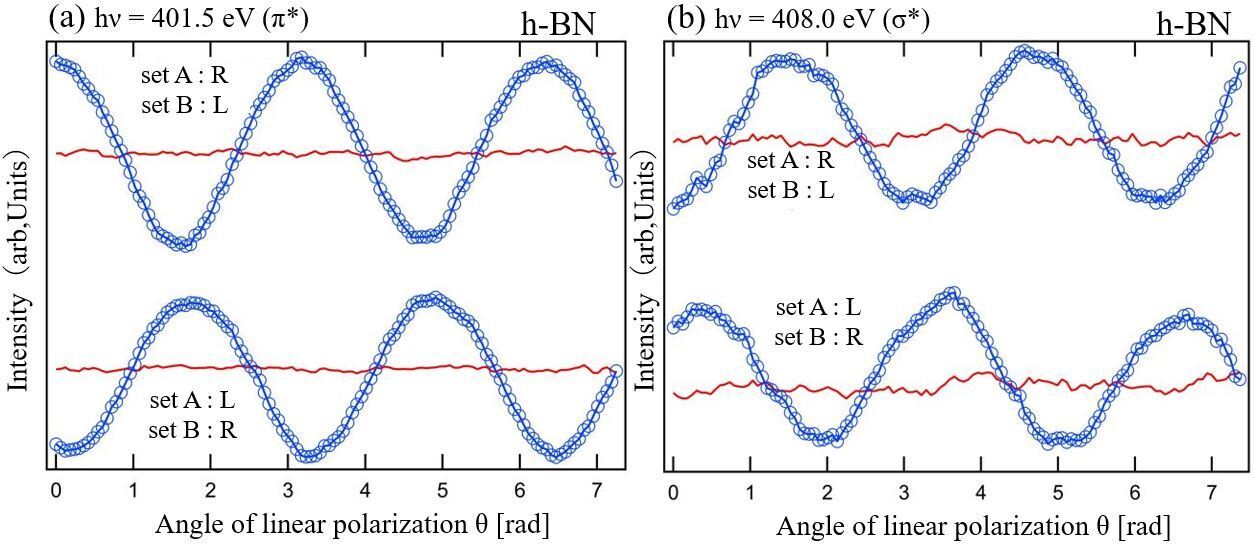}
   \caption{The rotational NEXAFS results of an oriented film of h-BN taken at incident angle of 80$^{\circ}$ (blue) and 0$^{\circ}$ (red). Angle ($\theta$) -dependence of intensity is taken at (a) hv=401.5 eV ($\pi^{*}$) and (b) at 408.0 eV ($\sigma^{*}$).} 
  \label{fig_3} 
 \end{center}  
\end{figure}

\begin{figure}[htp]
 \begin{center}
  \includegraphics[width=6cm]{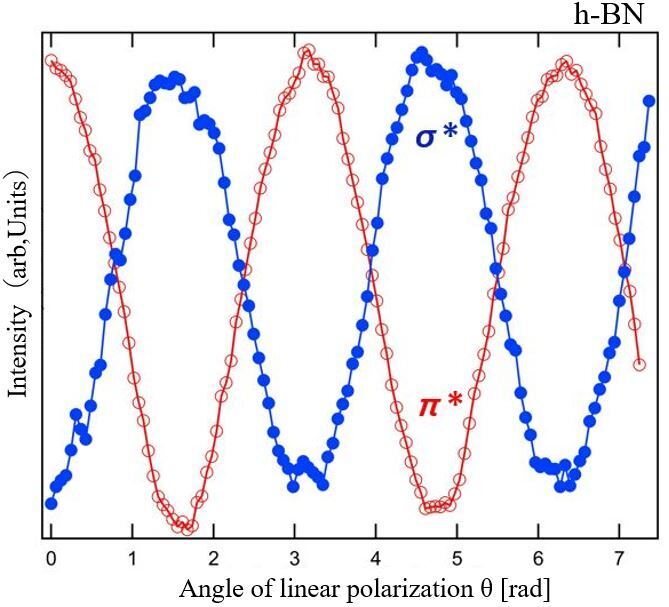}
   \caption{ An intensity vs angle plot of rotational NEXAFS for $\pi^{*}$ and $\sigma^{*}$ orbitals of the 2D h-BN film. The data were taken at the incident angle of 80$^{\circ}$.} 
  \label{fig_4} 
 \end{center}  
\end{figure}

Focusing on the $\pi^{*}$ and $\sigma^{*}$ peaks, we now trace their intensity variation with angle of the linear polarization, $\theta$, as shown in Fig.3. Since this NEXAFS methodology is different from the conventional, we hereafter call it the rotational NEXAFS measurement. When the incident angle is 80$^{\circ}$, intensity of hv=401.5 eV ($\pi^{*}$) becomes the maximum at $\theta$ =0, $\pi$, 2$\pi$ (horizontal) and the minimum at $\theta$ =$\pi$/2, 3$\pi$/2 (vertical), while that of hv=408.0 eV ($\sigma^{*}$) shows the opposite. When a combination of the circularly polarized light in set A and B is reversed, intensity of the individual orbitals is consistently reversed. Concerning the rotational NEXAFS data, taken at incident angle of 0$^{\circ}$, it shows no angle-dependence for both the $\pi^{*}$ and $\sigma^{*}$ peaks. These results of are consistent with those taken by the conventional NEXAFS, shown in Fig.2. However, a set of the rotational NEXAFS data is composed of more data points of the angle-dependence that increases the experimental reliability. The rotational NEXAFS experiment summarizes the clear electronic contrast between the $\pi^{*}$ (out-of-plane) and $\sigma^{*}$ (in-plane) orbital of h-BN, as shown in Fig.4. A style of the plot is expected to be useful in making assignments of the electronic states and in determining the precise configuration of molecules. 

A method of rotational NEXAFS is essentially spectroscopic applications of a segmented undulator. Polarization controls by interference of electromagnetic waves provides continuous variations of circular, elliptical, and linear polarization during soft X-ray experiments. In addition, adoption of electromagnetic coil in PS makes the fast switching that lead to quick measurement and high-sensitive signal detection by lock-in amplification techniques\cite{Kubota2017,Matsuda2014}. Since a segmented undulator provides the stable beam position on a sample during the polarization change, it has good compatibility with experiments of nanofocusing, imaging, or spectromicrosopy\cite{Zajac2021,Watt2012,Fukidome2014}. Furthermore, the additional quick and sensitive detections by fast switching benefits $operando$ experiments that captures a faint change in a non-uniform functional material such as catalysts or nanodevices. 

\section{Conclusion}
We developed a new method for NEXAFS experiment. It was performed by continuously rotating linearly polarized light, generated by a segmented undulator. We demonstrated the rotational NEXAFS experiment on a 2D film of h-BN and successfully obtained detailed datasets of the $\pi^{*}$ (out-of-plane) and $\sigma^{*}$ (in-plane) orbitals. The novel experimental promisingly provides a number of advantages as compared with the conventional one. 1) It significantly increases a number of angle-dependent datapoint that makes a precise determination of molecule structure. 2) Since it fixes the beam position during polarization controls, it benefits experiments of samples in nanospace or under the $operando$ condition. 3) It also paves a way to develop the high sensitive measurements using lock-in amplification when it is combined with fast polarization-switching.

\section*{Acknowledgement}
 This work was partially supported by Grant-in-Aid for Scientific Research (KAKENHI 18H03874, 19H04398) from the Japan Society for the Promotion of Science (JSPS) and the NewSUBARU SR facility at the University of Hyogo. The NEXAFS experiments were carried out as a research at the Synchrotron Radiation Research Organization, the University of Tokyo.

\section*{References}


\begin{thebibliography}{99}
{

\bibitem{Stohr} J. Stohr, \textit{NEXAFS Spectroscopy} (Springer, Berlin, 2003).

\bibitem{Zhang2020} J. L. Zhang, X. Ye, C. Gu, C. Han, S. Sun, L. Wang, W. Chen, Surf. Sci. Rep. \textbf{75}, 100481 (2020).
    https://doi.org/10.1016/j.surfrep.2020.100481
\bibitem{Kobl2020} J. Kobl, D. Wechsler, E. Y. Kataev, F. J. Williams, N. Tsud, S. Franchi, H.-P. Steinruck, O. Lytken, Surf. Sci. \textbf{698}, 121612 (2020).
    https://doi.org/10.1016/j.susc.2020.121612   

\bibitem{Tateishi} I. Tateishi, N. T. Cuong, C.A.S. Moura, M. Cameau, R. Ishibiki, A. Fujino, S. Okada, A. Yamamoto, M. Araki, S. Ito, S. Yamamoto, M. Niibe, T. Tokushima, D.E. Weibel, T. Kondo, M. Ogata, and I. Matsuda, Phys. Rev. Materials \textbf{3}, 024004 (2019).
    https://doi.org/10.1103/PhysRevMaterials.3.024004

\bibitem{Yamamoto2019} S. Yamamoto, H. S. Kato, A. Ueda, S. Yoshimoto, Y. Hirata, J. Miyawaki, K. Yamamoto, Y. Harada, H. Wadati, H. Mori, J. Yoshinobu, I. Matsuda, e-J. Surf. Sci. Nanotechol. \textbf{17}, 49 (2019).
    https://doi.org/10.1380/ejssnt.2019.49

\bibitem{Auwarter2018} W. Auwarter, Surf. Sci. Rep. \textbf{74}, 1 (2019).     
    https://doi.org/10.1016/j.surfrep.2018.10.001
\bibitem{Niibe2018} M. Niibe, N. Takehira, and Takashi Tokushima, e-J. Surf. Sci. Nanotech. \textbf{16}, 122 (2018). 
    https://doi.org/10.1380/ejssnt.2018.122
    
\bibitem{Zajac2021} M. Zajac, T. Giela, K. Freindl, K. Kollbek, J. Korecki, E. Madej, K. Pitala, A. Koziol-Rachwal, M. Sikora, N. Spiridis, J. Stepien, A. Szkudlarek, M. Slezak, T. Slezak, D. Wilgocka-Slezak, Nucl. Instrum. Meth. Phys. Res. B \textbf{492}, 43 (2021).
    https://doi.org/10.1016/j.nimb.2020.12.024
\bibitem{Watt2012}B. Watts and H. Ade, Materials Today, \textbf{15}, 148 (2012).
    https://doi.org/10.1016/S1369-7021(12)70068-8
\bibitem{Fukidome2014} H. Fukidome, M. Kotsugi, K. Nagashio, et al., Sci Rep \textbf{4}, 3713 (2014). 
     https://doi.org/10.1038/srep03713
     

\bibitem{Sasaki1993} S. Sasaki, K. Kakuno, T. Takada, T. Shimada, K. Yanagida, Y. Miyahara, Nucl. Instrum. Meth. Phys. Res. A \textbf{331}, 763 (1993). 
  https://doi.org/10.1016/0168-9002(93)90153-9    
\bibitem{Sasaki1994} S. Sasaki, Nucl. Instrum. Meth. Phys. Res. A \textbf{347}, 83 (1994).
    https://doi.org/10.1016/0168-9002(94)91859-7
    
\bibitem{Tanaka2002} T. Tanaka, H. Kitamura, Nucl. Instrum. Meth. Phys. Res. A 490, 583 (2002).
    https://doi.org/10.1016/S0168-9002(02)01094-X
\bibitem{Tanaka2004} T. Tanaka, H. Kitamura, AIP Conference Proceedings \textbf{705}, 231 (2004).
    https://doi.org/10.1063/1.1757776

\bibitem{Yamamoto2014} S. Yamamoto, Y. Senba, T. Tanaka, H. Ohashi, T. Hirono, H. Kimura, M. Fujisawa, J. Miyawaki, A. Harasawa, T. Seike, S. Takahashi, N. Nariyama, T. Matsushita, M. Takeuchi, T. Ohata, Y. Furukawa, K. Takeshita, S. Goto, Y. Harada, S. Shin, H. Kitamura, A. Kakizaki, M. Oshima and I. Matsuda, J. Synchrotron Rad. \textbf{21}, 352 (2014). 
    https://doi.org/10.1107/S1600577513034796
\bibitem{Matsuda2019} I. Matsuda, S. Yamamoto, J. Miyawaki, T. Abukawa, and T. Tanaka, e-J. Surf. Sci. Nanotechnol. \textbf{17}, 41 (2019).
    https://doi.org/10.1380/ejssnt.2019.41
\bibitem{Kubota2017} Y. Kubota, Y. Hirata, J. Miyawaki, S. Yamamoto, H. Akai, R. Hobara, Sh. Yamamoto, K. Yamamoto, T. Someya, K. Takubo, Y. Yokoyama, M. Araki, M. Taguchi, Y. Harada, H. Wadati, M. Tsunoda, R. Kinjo, A. Kagamihata, T. Seike, M. Takeuchi, T. Tanaka, S. Shin, and I. Matsuda, Phys. Rev. B \textbf{96}, 214417 (2017).
    https://doi.org/10.1103/PhysRevB.96.214417
 \bibitem{Senba2011} Y. Senba, S. Yamamoto, H. Ohashi, I. Matsuda, M. Fujisawa, A. Harasawa, T. Okuda, S. Takahashi, N. Nariyama, T. Matsushita, T. Ohata, Y. Furukawa, T. Tanaka, K. Takeshita, S. Goto, H. Kitamura, A. Kakizaki and M. Oshima, Nucl. Instr. Meth. Phys. Res. A \textbf{649}, 58 (2011).   
     https://doi.org/10.1016/j.nima.2010.12.242

\bibitem{Matsuda2014} I. Matsuda, A. Kuroda, J. Miyawaki, Y. Kosegawa, S. Yamamoto, T. Seike, T. Bizen, Y. Harada, T. Tanaka, and H. Kitamura, Nucl. Instrum. Meth. Phys. Res. A \textbf{767}, 296 (2014).
    https://doi.org/10.1016/j.nima.2014.08.037
    }
\end{thebibliography}

\end{document}